# EFFECT OF THE SEXTUPOLE FINITE LENGTH ON DYNAMIC APERTURE IN THE COLLIDER FINAL FOCUS

A. Bogomyagkov, S. Glukhov, E. Levichev and P. Piminov
*Budker Institute of Nuclear Physics, Novosibirsk 630090, Russia*

*Abstract*

In a collider final focus (FF) system chromaticity generated by FF quadrupoles is compensated by sextupole pairs with *–I* map in between to cancel geometrical aberrations. However, it works perfectly only for zero-length sextupole kicks; for realistic thick magnets only quadratic aberrations are cancelled exactly while higher order terms appear and may deteriorate the transverse dynamic aperture. The latter is particularly important for prospective projects of the future circular colliders based on the Crab-Waist collision approach where extremely low beta at IP generates very large chromatic effects. In the present paper, we study these high aberrations by the power series technique and propose the method of the dynamic aperture recover. Results of analytic study are compared with those obtained by tracking.

## 1. INTRODUCTION

Collider's final focus (FF) system serves for transverse squeezing of the beam in the interaction point (IP) and thus providing high luminosity. Chromaticity generated by FF lens is estimated as

$$\xi^* \approx -l/\beta^*,$$

where $l$ is a distance to the closest lens and $\beta^*$ is a beta function value at IP. Colliders with Crab-Waist (CW) [1, 2] interaction scheme possess $\beta_y^* < 1$ mm and vertical chromaticity $\xi_y^* \sim -10^3 \ldots -10^4$. Horizontal chromaticity is smaller in case of flat beams, however its value is still significant — several hundreds. The major part of the ring chromaticity belongs to the FF section.

The local chromaticity compensation scheme (of betatron frequencies and beta functions) utilizes sextupoles, which due to high values of the chromaticity appear to be rather strong and therefore significantly degrade dynamic aperture. In order to avoid degradation, pairs of sextupoles are used, with *–I* transformation inside the pair Fig.1.1. The second sextupole cancels nonlinear aberrations of the first one and the whole system does not disturb the beam dynamics.

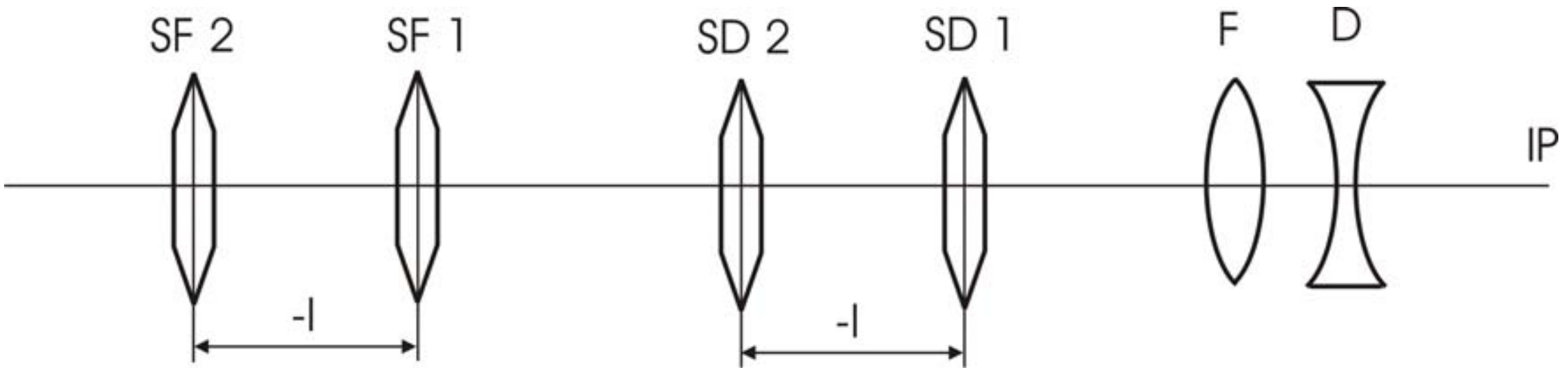

Fig.1.1 Chromaticity compensation by sextupoles pairs

However, simulation and analytical studies show the full cancelation is possible only for zero length sextupoles (so-called *kick approximation*): $B''(s) = B''(s_0) \cdot L \cdot \delta(s - s_0)$. Considera-tion of the sextupoles with realistic lengths and separated by *–I* map shows that full cancelation does not happen and dynamic aperture is again small.

The present paper discusses reasons for dynamic aperture deterioration due to non-zero length sextupoles, gives an analytical solution describing particles propagation through "thick" sextupole in the form of power series, presents ways to minimize nonlinear aberrations and increase dynamic aperture.

## 2. DYNAMIC APERTURE VS MULTIPOLE LENGTH

In order to figure out the order of the nonlinear aberrations mostly affecting dynamic aperture, let us study dependence of the stability area size on multipole magnet length.

We start with sextupole magnet. Hamiltonian of sextupole perturbation is

$$H_1 = k_2(s)\left(x^3 - 3xy^2\right)/6, \tag{2.1}$$

where $k_2(s) = B_y''/(cp)$. The standard transformation to variables "action-phase" $(J,\varphi)$ and usage of perturbation theory allows obtaining of new action variable $\bar{J}$ – invariant of motion in second order infinitesimal [3]:

$$\bar{J}_x = J_x + 3\sqrt{8}J_x^{3/2}\left[\cos\varphi_x \sum_n \frac{A_{1n}}{\nu_x - n} + \cos 3\varphi_x \sum_n \frac{A_{3n}}{3\nu_x - n}\right] = const, \tag{2.2}$$

where, for simplicity, only horizontal motion is presented, perturbation is assumed to be even, and a constant $\bar{J}_x$ is found from initial conditions. Fourier harmonics of perturbation potential are found as ($j = 1,3$)

$$A_{jn} = \frac{1}{48\pi}\sum_m \beta_{xm}^{3/2}(k_2L)_m \cos(j\psi_x - \nu_x\theta + n\theta)_m, \tag{2.3}$$

where summation is performed over all sextupoles with integrated strength $(k_2L)_m$, $\theta = s/\bar{R}$ is azimuth angle, $\nu_x$ and $\psi_x(s)$ are betatron frequency and phase respectfully.

Equation (2.2) defines phase trajectories $J_x(\varphi_x)$ with accuracy up to second order, as it shown on Fig.2.1. We will define the border of stability area as break point of $J_x(\varphi_x)$ curve: $dJ_x/d\varphi_x = \infty$ and we will write the full derivative of (2.2) in form

$$\frac{d\bar{J}_x}{d\varphi_x} = \frac{\partial\bar{J}_x}{\partial\varphi_x} + \frac{\partial\bar{J}_x}{\partial J_x}\frac{\partial J_x}{\partial\varphi_x} = 0, \qquad \text{or} \qquad \frac{\partial J_x}{\partial\varphi_x} = \infty = -\frac{\partial\bar{J}_x/\partial\varphi_x}{\partial\bar{J}_x/\partial J_x},$$

from which we derive

$$\bar{J}_x' = \partial\bar{J}_x/\partial J_x = 0. \tag{2.4}$$

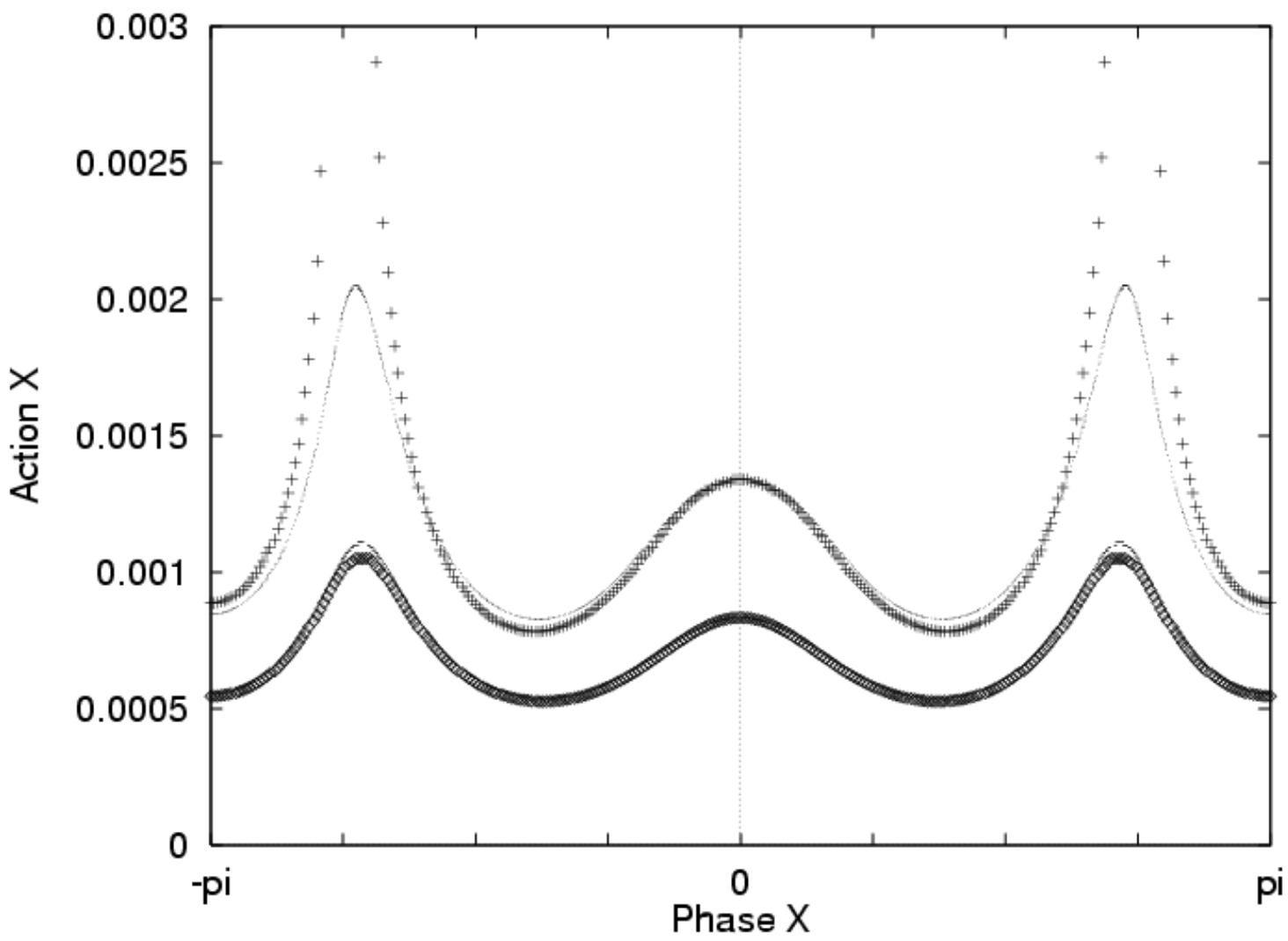

Fig.2.1 Stable and unstable phase trajectories found with the help of (2.2)

Solution of the last equation gives an estimate of the stability area size $J_{x\lim} = F(\varphi_{x\lim})$. Since obtained expression is a function of the phase variable, it is necessary to find such a value $\varphi_{x\lim}$, which gives $J_{x\lim} = \min$

$$\partial J_{x\lim} / \partial\varphi_x = 0 . \tag{2.5}$$

It is interesting to note that in vicinity of the resonance (e.g. integer) where $\nu_x - n = \delta << 1$, and neglecting non-resonance terms in (2.2) one obtains

$$\bar{J}_x \approx J_x + 3\sqrt{8} J_x^{3/2} \frac{A_{1n}}{\nu_x - n}\cos\varphi_x, \quad \text{или} \quad H_r = \delta\cdot\bar{J}_x \approx \delta\cdot J_x + 3\sqrt{8}J_x^{3/2} A_{1n}\cos\varphi_x . \tag{2.6}$$

The last expression is exactly resonance Hamiltonian, obtained with canonical transformations into the rotating coordinate system, and conditions (2.4) and (2.5) are equations for turning points, which define border of the stability area in vicinity of isolated resonance.

Thus, usage of action invariant, found with the help of non-resonant theory of perturbations, for finding borders of stability area, could be used in the whole frequency domain as in vicinity of the resonance and in far off. The method could be expanded for the case of 2D motion and for the higher orders of perturbation theory.

Applying condition (2.4) to (2.2), we obtain an estimate of the dynamic aperture

$$A_{x\lim} \sim J_{x\lim}^{1/2} = \left\{-9\sqrt{2}\left[\cos\varphi_x \sum_n \frac{A_{1n}}{\nu_x - n} + \cos 3\varphi_x \sum_n \frac{A_{3n}}{3\nu_x - n}\right]\right\}^{-1} \sim L^{-1}, \tag{2.7}$$

i.e. dynamic aperture determined by quadratic (sextupole) nonlinearity decreases as $1/L$ (with accuracy up to second order infinitesimal) with increase of the sextupole length.

The search of $\varphi_{x\lim}$, which gives $J_{x\lim}(\varphi_{x\lim}) = \min$, results in numerical value of dynamic aperture for particular values of perturbation harmonics, however, it exceeds the scope of the present paper.

For the cubic nonlinearity (here it is octupole), Hamiltonian is

$$H_2 = k_3(s)\left(x^4 - 6x^2y^2 + y^2\right)/24 , \tag{2.8}$$

where $k_3(s) = B_y'''/(cp)$, and second order invariant (2.2) is

$$\bar{J}_x = J_x + 16 \cdot J_x^2 \left[ \cos 4\varphi_x \sum_n \frac{A_{4n}}{4\nu_x - n} + \cos 2\varphi_x \sum_n \frac{2A_{2n}}{2\nu_x - n} \right] = const, \qquad (2.9)$$

where perturbation harmonics ( $j = 2,4$ ) are equal

$$A_{jn} = \frac{1}{384\pi} \sum_m \beta_{xm}^2 (k_3 L)_m \cos(j\psi_x - \nu_x \theta + n\theta)_m . \qquad (2.10)$$

Applying (2.4) to (2.8) we obtain

$$A_{x\lim}^2 \sim J_{x\lim} \sim L^{-1},$$

i.e. for cubic nonlinearity dynamic aperture decreases as $A_x \sim L^{-1/2}$.

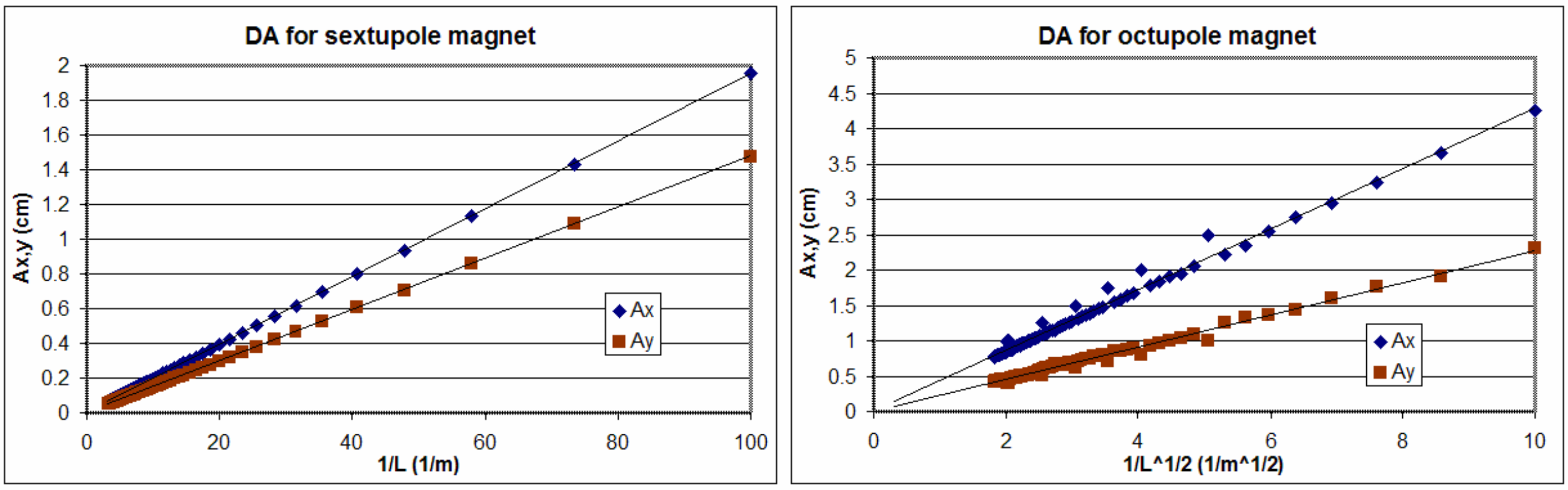

Fig.2.2 DA vs. the magnet length for sextupole (left, $\sim L^{-1}$) and octupole (right, $\sim L^{-1/2}$)

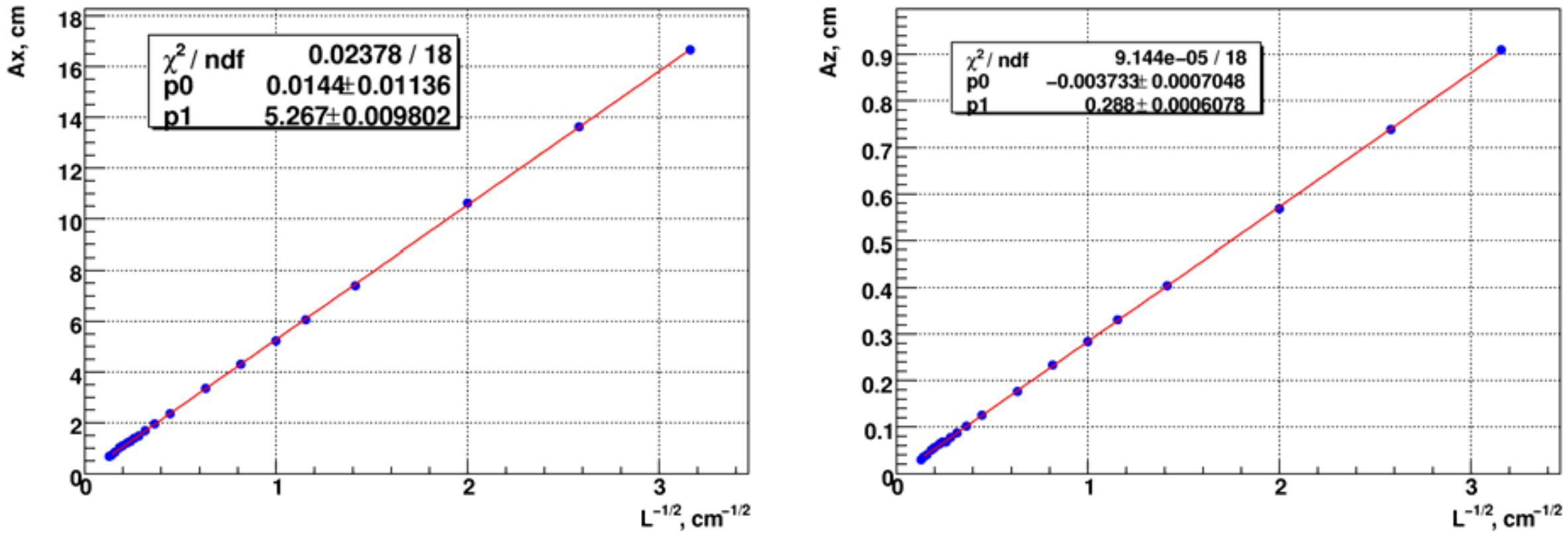

Fig.2.3 DA vs. the magnet length for the sextupole pair with $-I$ map in between (left – horizontal, right – vertical)

Results of the simulation for discussed above sextupole and octupole perturbation are shown on Fig.2.2. As one can see the law of the aperture decrease $A_{xs} \sim L^{-1}$ and $A_{xo} \sim L^{-1/2}$ is sufficiently fulfilled for both transverse coordinates in all domain of betatron frequencies.

Simulation performed for pair of sextupoles with $-I$ map shows that with increase of the sextupole length (keeping the integrated sextupole strength constant) dynamic aperture falls as $A_{x,y} \sim L^{-1/2}$ (Fig.2.3). This fact witnesses in favor of the third order nonlinearities being the source of dynamic aperture deterioration (not necessarily octupole). The law $A_{x,y} \sim L^{-1/2}$ is fulfilled with sufficient accuracy up to the border of the aperture (large amplitudes) hence aberrations of higher or lower orders are not significant.

## 3. POWER SERIES SOLUTION

In order to obtain aberrations terms of higher orders for particle propagating through multipole magnet with non-zero length, let us search for solution of corresponding equations in the form of power series of longitudinal variable *s*:

$$x(s) = \sum_{i=0}^{\infty} a_i \frac{s^i}{i!}, \qquad y(s) = \sum_{i=0}^{\infty} b_i \frac{s^i}{i!}. \tag{3.1}$$

Substitution of these expressions in equations for multipole magnets allows finding coefficients in the form of recursive expressions starting from

$$a_0 = x(0), \quad a_1 = x'(0), \quad b_0 = y(0), \quad b_1 = y'(0).$$

However, before proceeding to examples let us give some notes. Solution (3.1) is similar to one proposed earlier with help of Lie operators [4, 5]:

$$x(s) = \sum \frac{s^i}{i!} \frac{d^i x}{ds^i}\bigg|_{s=0} = \exp(s : -H :)x\Big|_{x(0),x'(0)}, \tag{3.2}$$

where $H$ is Hamiltonian, and a colon represents Poisson brackets. Knowing Hamiltonian and applying exponential Lie operators on vector of initial coordinates one is able to find solution at the end of the section (transformation). Continuous element is described by “sandwich” of “thin” multipoles with Hamiltonian $H(s_n)$ and linear sections with the length of $ds_n$ and transport matrix $M_n$ [5]:

$$\prod_{n=1}^{N} M_n \exp[ds_n : -H(s_n) :].$$

Hamiltonian of transformation through “thick” element is obtained by letting $ds \to 0$ and integrating of corresponding expressions along the element length.

Approach offered in our paper looks simpler and more convenient the Lie technique, because it does not require finding of exponential operators, recursive expressions immediately give aberration terms of higher orders. For description of continuous elements, there is no need to artificially divide elements into slices framed by linear transport sections with consequent integration along the longitudinal coordinate.

### *3.1 Quadrupole*

Equation of motion for quadrupole lens is

$$x'' = -k_1 \cdot x, \qquad y'' = k_1 \cdot y, \tag{3.1.1}$$

where $k_1(s) = B'_y/(cp)$. Substituting (3.1) into (3.1.1) and noting that

$$p_x = x'(s) = \sum_{i=0}^{\infty} a_{i+1} \frac{s^i}{i!}, \qquad x''(s) = \sum_{i=0}^{\infty} a_{i+2} \frac{s^i}{i!}, \tag{3.1.2}$$

(similar for $y(s)$) we obtain recursive expressions for coefficients:

$$a_{i+2} = -k_1 a_i, \qquad b_{i+2} = k_1 b_i. \tag{3.1.3}$$

Hence, for horizontal motion we can write

$$\begin{aligned} a_0 &= x_0, & a_1 &= x'_0 = p_{x0}, \\ a_2 &= -k_1 x_0, & a_3 &= -k_1 x'_0, \\ a_4 &= k_1^2 x_0, & a_5 &= k_1^2 x'_0 \end{aligned}$$

and so on, where $x_0$ and $x'_0$ are initial conditions. Substitution of obtained coefficients into (3.1) yields

$$x(s) = x_0\left(1 - k_1\frac{s^2}{2!} + k_1^2\frac{s^4}{4!} - ...\right) + x'_0\left(s - k_1\frac{s^3}{3!} + k_1^2\frac{s^5}{5!} - ...\right) = x_0 \cos\left(\sqrt{k_1}s\right) + \frac{x'_0}{\sqrt{k_1}}\sin\left(\sqrt{k_1}s\right).$$

Similar procedure could be performed for $y(s)$, and one can see that series tend to solutions known from the theory of linear betatron oscillations.

### *3.2 Sextupole*

Equation of motion for sextupole lens is

$$x'' = -\frac{k_2}{2}\left(x^2 - y^2\right), \qquad y'' = k_2 \cdot xy, \tag{3.2.1}$$

where $k_2(s) = B''_y/(cp)$. Substitution of (3.1) into (3.2.1) with the help of Cauchy formula

$$\left(\sum_{i=0}^{\infty} a_i\right)\left(\sum_{i=0}^{\infty} b_i\right) = \sum_{i=0}^{\infty}\sum_{j=0}^{i} a_i b_{i-j}$$

gives recursive expressions for coefficients

$$a_{i+2} = -\frac{k_2}{2}\sum_{j=0}^{i} C_i^j\left(a_j a_{i-j} - b_j b_{i-j}\right), \qquad b_{i+2} = k_2 \sum_{j=0}^{i} C_i^j a_j b_{i-j}, \tag{3.2.2}$$

where $C_i^j = i!/j!(i-j)!$. Using (3.2.2) and initial conditions at $s = 0$ we obtain

$$a_0 = x_0 \qquad\qquad a_1 = x'_0 = p_{x0},$$

$$a_2 = -\frac{k_2}{2}\left(x_0^2 - y_0^2\right), \qquad a_3 = k_2\left(-p_{x0}x_0 + p_{y0}y_0\right),$$

$$a_4 = \frac{k_2}{2}\left(-2p_{x0}^2 + 2p_{y0}^2 + k_2x_0\left(x_0^2 + y_0^2\right)\right), \qquad a_5 = \frac{k_2^2}{2}\left(5p_{x0}x_0^2 + 6p_{y0}x_0y_0 - p_{x0}y_0^2\right),$$

$$a_6 = -\frac{k_2^2}{4}\left(-20p_{x0}^2x_0 - 12p_{y0}^2x_0 - 8p_{x0}p_{y0}y_0 + k_2\left(5x_0^4 - 18x_0^2y_0^2 + y_0^4\right)\right),$$

$$a_7 = k_2^2\left(5p_{x0}^3 + 5p_{x0}p_{y0}^2 + 14k_2p_{y0}x_0^2y_0 + 2k_2x_0p_{x0}\left(-5x_0^2 + 8y_0^2\right)\right),$$

$$a_8 = \frac{k_2^3}{2}\left(140p_{x0}p_{y0}x_0y_0 + p_{y0}^2\left(23x_0^2 + 5y_0^2\right) + p_{x0}^2\left(-75x_0^2 + 47y_0^2\right) + 2k_2x_0\left(5x_0^4 + x_0^2y_0^2 + 8y_0^4\right)\right).$$

$$b_0 = y_0 \qquad b_1 = y_0' = p_{y0},$$

$$b_2 = k_2x_0y_0, \qquad b_3 = k_2\left(p_{x0}x_0 + p_{y0}y_0\right),$$

$$b_4 = \frac{k_2}{2}\left(4p_{x0}p_{y0} + k_2y_0\left(x_0^2 + y_0^2\right)\right), \qquad b_5 = \frac{k_2^2}{2}\left(-p_{y0}x_0^2 + 6p_{x0}x_0y_0 + 5p_{y0}y_0^2\right),$$

$$b_6 = k_2^2\left(2p_{x0}p_{y0}x_0 + 3p_{x0}^2y_0 + 5p_{y0}^2y_0 - 2k_2x_0y_0\left(x_0^2 - 2y_0^2\right)\right),$$

$$b_7 = k_2^2\left(5p_{x0}^2p_{y0} - 7k_2p_{x0}y_0\left(x_0^2 - y_0^2\right) + p_{y0}\left(5p_{y0}^2 - 3k_2x_0^3 + 23k_2x_0y_0^2\right)\right),$$

$$b_8 = \frac{k_2^3}{2}\left(-18p_{x0}^2x_0y_0 + p_{x0}p_{y0}\left(-42x_0^2 + 98y_0^2\right) + y_0\left(122p_{y0}^2x_0 + k_2\left(x_0^4 + 32x_0^2y_0^2 + 7y_0^4\right)\right)\right).$$

As one can see, monomials of third order in initial coordinates appear in the first eight coefficients. Thus, it is possible to describe particle motion in thick sextupole with accuracy up to third order by the summation of the first eight monomials (order of $L^7$).

### *3.3 Octupole*

Equation of motion for octupole is

$$x'' = -\frac{k_3}{6}\left(x^3 - 3xy^2\right), \qquad y'' = \frac{k_3}{6}\left(3x^2y - y^3\right), \tag{3.3.1}$$

where $k_3(s) = B_y'''/(cp)$. Obtained recursive expressions for coefficients are

$$\begin{aligned} a_{k+2} &= -\frac{k_3}{6}\sum_{i=0}^{k}\sum_{j=0}^{k-i} C_k^i C_{k-i}^j \left(a_i a_j a_{k-i-j} - 3a_i b_j b_{k-i-j}\right), \\ b_{k+2} &= \frac{k_3}{6}\sum_{i=0}^{k}\sum_{j=0}^{k-i} C_k^i C_{k-i}^j \left(3b_i a_j a_{k-i-j} - b_i b_j b_{k-i-j}\right), \end{aligned} \tag{3.3.2}$$

The first coefficients of solution (3.1) are:

$$a_0 = x_0 \qquad a_1 = x_0' = p_{x0},$$

$$a_2 = -\frac{k_3}{6}\left(x_0^3 - 3x_0y_0^2\right), \qquad a_3 = \frac{k_3}{2}\left(2p_{y0}x_0y_0 + p_{x0}\left(y_0^2 - x_0^2\right)\right),$$

$$a_4 = \frac{k_3}{12}\left(-12p_{x0}^2x_0 + 24p_{x0}p_{yo}y_0 + x_0\left(12p_{y0}^2 + k_3\left(x_0^2 + y_0^2\right)^2\right)\right),$$

$$b_0 = y_0 \qquad b_1 = y_0' = p_{y0},$$

$$b_2 = \frac{k_3}{6}\left(3x_0^2 y_0 - y_0^3\right), \qquad b_3 = \frac{k_3}{2}\left(2p_{y0}x_0 y_0 + p_{y0}\left(x_0^2 - y_0^2\right)\right),$$

$$b_4 = \frac{k_3}{12}\left(24 p_{x0} p_{y0} x_0 + 12 p_{x0}^2 y_0 - x_0\left(12 p_{y0}^2 - k_3\left(x_0^2 + y_0^2\right)^2\right)\right),$$

## 4. CORRECTION OF THE LENGTH EFFECTS FOR THE SEXTUPOLES WITH –I TRANSFORMATION IN BETWEEN

### *4.1 Sextupole pair*

Layout of sextupole placement is shown on Fig.4.1.1. The length $L$ and sextupole strengths $k_2(s) = B_y''/(cp)$ are the same.

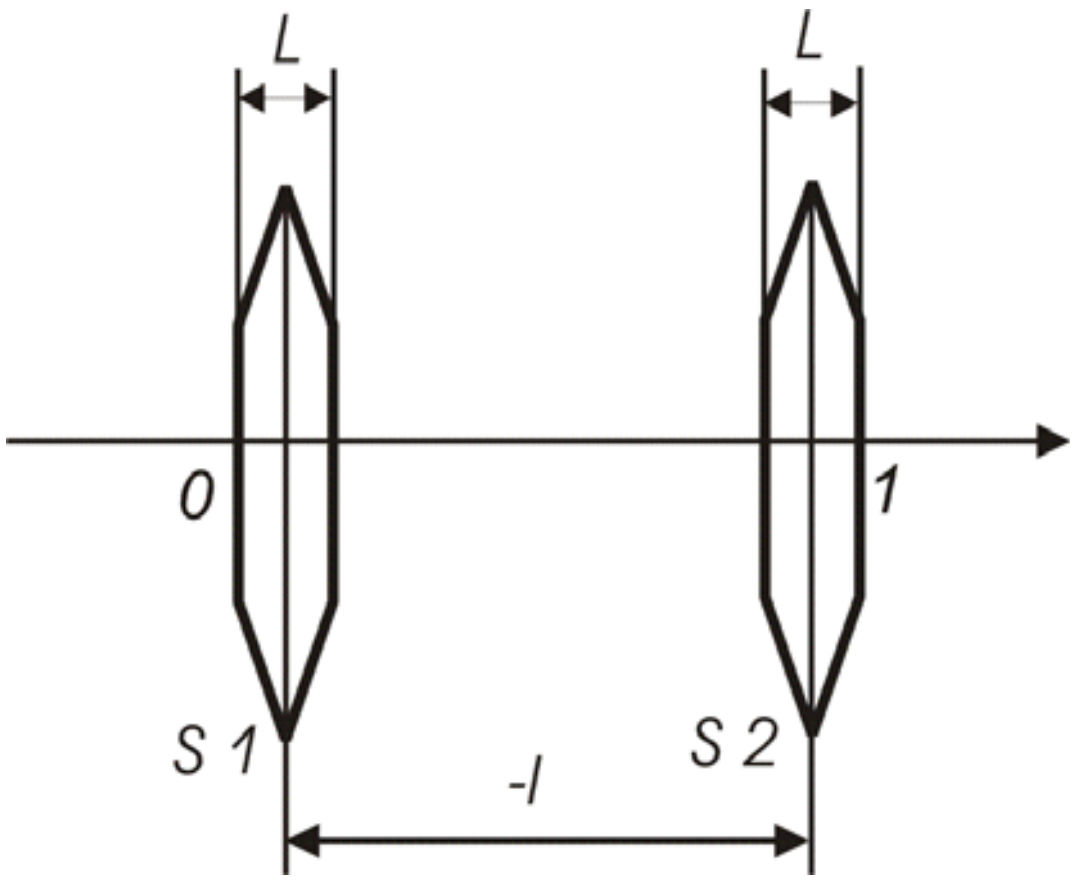

Fig.4.1.1 Pair of sextupoles with the same length and strength with –*I* map between the centers

With accuracy up to third order in initial coordinates transformation from point 0 to point 1 (Fig.4.1.1) is

$$x_1 = -x_0 - p_{x0}L - \frac{(k_2 L)^2}{12}\left(x_0^3 + x_0 y_0^2\right)L^2 + O(L^5),$$

$$p_{x1} = -p_{x0} - \frac{(k_2 L)^2}{6}\left(x_0^3 + x_0 y_0^2\right)L - \frac{(k_2 L)^2}{12}\left(3p_{x0}x_0^2 + 2p_{y0}x_0 y_0 + p_{x0}y_0^2\right)L^2 + O(L^5),$$

$$y_1 = -y_0 - p_{y0}L - \frac{(k_2 L)^2}{12}\left(x_0^2 y_0 + y_0^3\right)L^2 + O(L^5), \tag{4.1.1}$$

$$p_{y1} = -p_{y0} - \frac{(k_2 L)^2}{6}\left(x_0^2 y_0 + y_0^3\right)L - \frac{(k_2 L)^2}{12}\left(p_{y0}x_0^2 + 2p_{x0}x_0 y_0 + 3p_{y0}y_0^2\right)L^2 + O(L^5).$$

Note, that cubic monomial of obtained expression does not coincide with octupole (e.g. $x_1 = x_0 + p_{x0}L - \frac{k_3}{12}\left(x_0^3 - 3x_0 y_0^2\right)L^2$) and thus it is not possible to cancel this monomial by octupole lenses. The value of cubic perturbation for continuous sextupole is integrated strength squared $\left(k_2 L\right)^2$.

*4.2 Correction of the sextupole length effect*

Let us try to compensate, for pair of sextupoles with –*I* map in between, effects of third order perturbation, caused by non-zero length of the sextupoles. For that, let us add the second sextupole pair as it shown on Fig.4.2.1. The integrated sextupoles strengths we will denote as $S_1 = k_{21}L$ and $S_2 = k_{22}L$. The distance between sextupoles of different pairs, we will express as $\Delta L = kL$.

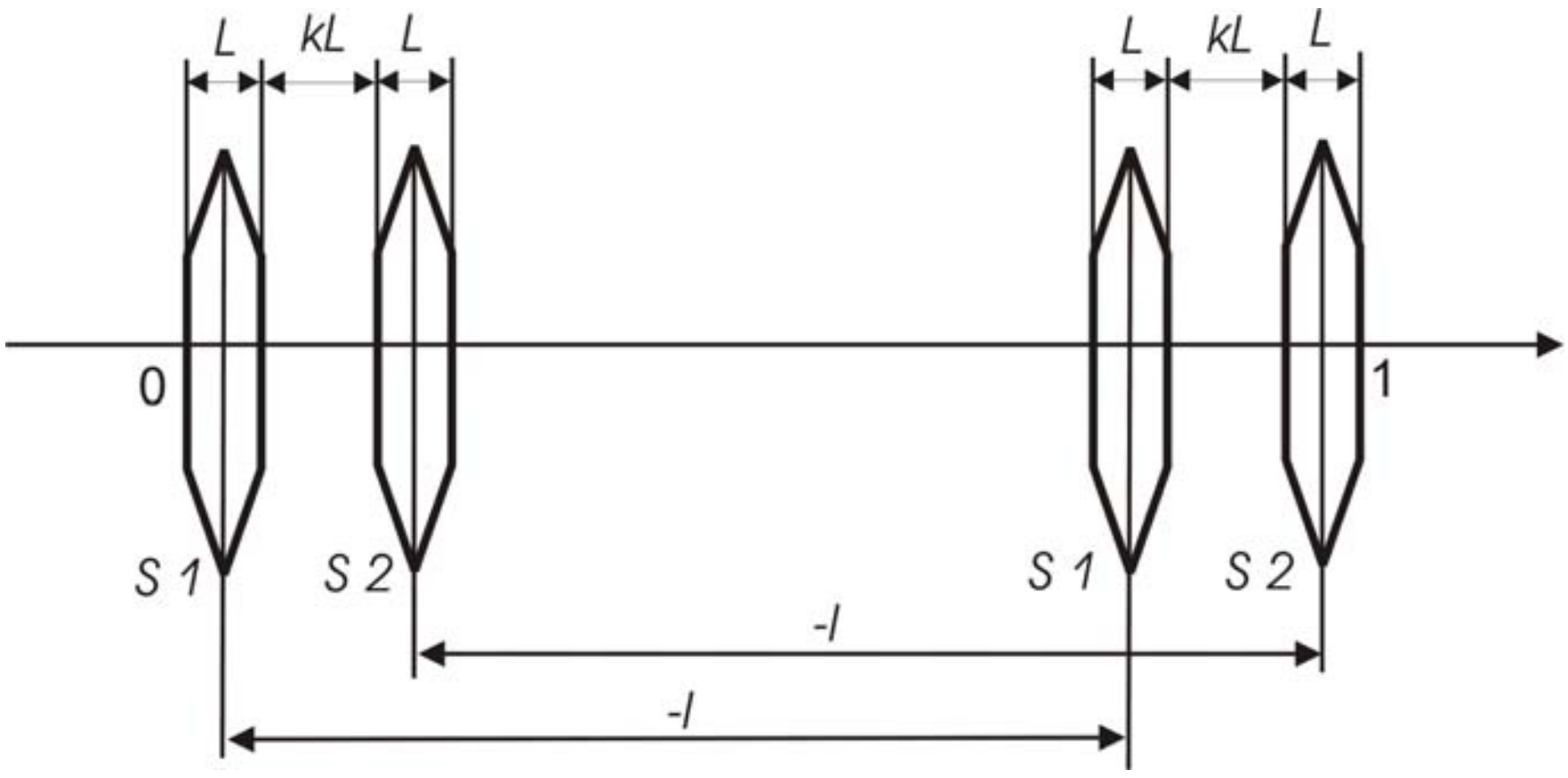

Fig.4.2.1 Two pairs of sextupoles

Let us write transformation between points 0 and 1. Since this transformation contains large number of monomials, we will limit ourselves to the third order of initial coordinates and will simplify initial conditions that $p_{x0} = p_{y0} = 0$. The reasons for such a simplification are that in chromaticity correction section corresponding betatron functions are large and transverse momentums $p_x \sim x/\beta_x$, $p_y \sim y/\beta_y$ are suppressed. With respect of what has been noted, the transformation is

$$x_1 = -x_0 - \frac{B_1(S_1,S_2,k)}{12}\left(x_0^3 + x_0 y_0^2\right)L^4, \qquad p_{x1} = -\frac{B_2(S_1,S_2k)}{6}\left(x_0^3 + x_0 y_0^2\right)L^3, \quad (4.2.1)$$

$$y_1 = -y_0 - \frac{B_1(S_1,S_2,k)}{12}\left(y_0^3 + x_0^2 y_0\right)L^4, \qquad p_{y1} = -\frac{B_2(S_1,S_2k)}{6}\left(y_0^3 + x_0^2 y_0\right)L^3,$$

$$B_1(S_1,S_2,k) = (3+2k)S_1^2 + 6\left(2+3k+k^2\right)S_1S_2 + S_2^2, \quad (4.2.2)$$

$$B_2(S_1,S_2,k) = S_1^2 + 6(1+k)S_1S_2 + S_2^2. \quad (4.2.3)$$

As one can see, coefficient $B_1(S_1,S_2,k)$ is responsible for aberrations in coordinate part of transformation, and $B_2(S_1,S_2,k)$ – in momentum part.

In order to restore –*I*, it is necessary to choose such strengths of correcting sextupoles and theirs position with respect to main ones (parameter $k = \Delta L/L$) which will provide $B_1 = B_2 = 0$. It is pity that quadratic equations (4.2.2) and (4.2.3) do not posses common roots, therefore it makes to consider separately conditions for cancelations of $B_1$ and $B_2$. However, there are two possible solutions depending on which pair is main one and which is correcting.

Let us start from the $S_1$ being the main pair, and $S_2$ being correcting.

Solution of quadratic equation (4.2.2) with respect to $S_2$ allows finding ratio $S_2/S_1$ as a function of distance between the main sextupole and correcting one (in terms of $k = \Delta L/L$). Do-ing that we cancel coordinate part of aberrations ($B_1 = 0$). Substituting obtained solution into $B_2$ allows figuring out the change in momentum part of transformation shown on Fig. 4.2.1.

Then procedure is repeated for equation (4.2.3), canceling momentum nonlinear part $B_2 = 0$ of transformation (4.2.1). Substitution of derived solution gives function for $B_1$.

Results of corresponding case are shown on Fig. 4.2.2. The left plot shows the relative strength of correcting sextupole $S_2$, necessary to cancel coefficient $B_1$ or $B_2$. The right plot presents the value of remained not canceled coefficient relative to its value when correcting sextupole is turned off ($S_2 = 0$).

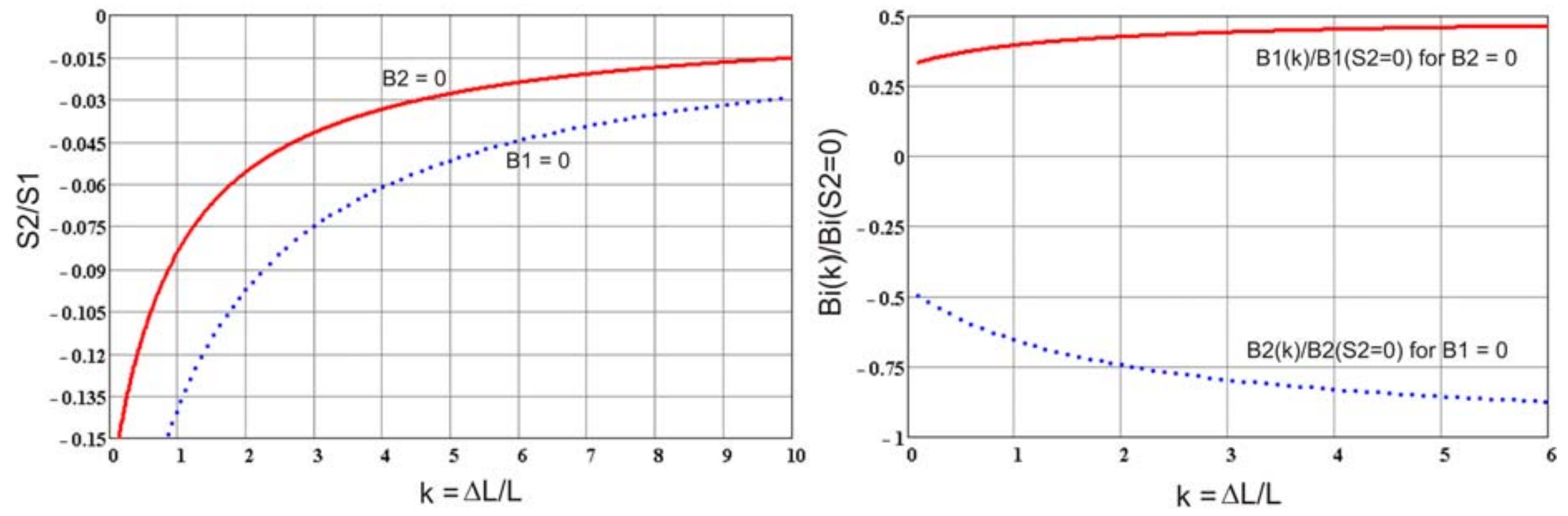

Fig.4.2.2 The ratio of correcting and main sextupole strengths (left) and relative decrease of corresponding coefficient $B_{1,2}$ (right) as a function of distance between the main sextupole and correcting one

Let us discuss results shown on Fig.4.2.2. Correcting sextupole with strength of $S_2 = -0.139 S_1$, is placed at distance $\Delta L = L$ (where $L$ is sextupole length) from the main sextupole $S_1$, cancels coefficient $B_1 = 0$ (point $k = 1$ on dashed curve of the left plot Fig.4.2.2). At that coefficient $B_2$ is –0.65 from the initial value when correcting sextupole is turned off $S_2 = 0$ (dashed curve of the right plot Fig.4.2.2). In order to cancel coefficient $B_2$, the strength of correcting sextupole should be $S_2 = -0.084 S_1$. At that value of $B_1$ is 0.39 from initial value. Note that for both cases sign of the correcting sextupoles is opposite to the sign of the main ones.

In general case, we may say that usage of correcting sextupole pair as it shown on Fig.4.2.1, allows partially cancel and partially decrease values of third order aberrations of transformation (4.1.1), caused by non-zero sextupole lengths.

Similar results could be obtained for the case when $S_2$ is a main pair, and $S_1$ – correcting. The strengths of correcting sextupoles in this case are ~3-10% from the main ones, with signs opposite to the main ones.

*4.3 Chromatic aberrations*

Although obtaining large dynamic aperture for particle with momentum deviation $\delta = \Delta p / p_0$ is not the main goal of the present work, we will make note on this subject. For particles with $\delta \neq 0$, optics of the section between sextupoles does not provide *–I* map due to chromaticity of the betatron phase advance. Therefore (a) for these particles second order geometrical aberrations (chromo-geometrical) are generated and (b) there is direct oscillation frequency dependence on momentum deviation $\nu(\delta)$, thus the frequency could be shifted to strong resonance where particle motion will become unstable.

Optimization of chromatic effects requires particular knowledge of position and parameters of the magnetic elements which provide *–I* transformation.

Let us write the standard Hamiltonian for magnetic elements considering momentum deviation:

$$H = -(1+hx)\sqrt{1-p_x^2-p_y^2} - \frac{e}{pc}A_s(x,y,s), \qquad (4.3.1)$$

where $h = 1/\rho$ is orbit curvature. Let us expand the square root in power series neglecting monomials $p_x^4, p_x^2 p_y^2, p_y^4$ and higher order. Then we will substitute known expression for vector potential. Particles deviation from the designed orbit we will represent in usual form $x = x_\beta + \eta\delta$, where dispersion could be written as Taylor series $\eta = \eta_0 + \eta_1\delta + \ldots$. After that we obtain the following expressions describing betatron oscillations and evolution of the dispersion function with accuracy up to second order of momentum deviation $\delta^2$.

*Dipole* (neglected monomials with $h^3, p_x^2, p_y^2, xp_x^2, x\delta^2, x^2\delta$).

$$x_\beta'' + h^2\left(1-2\delta+\delta^2\right)\cdot x_\beta = 0, \qquad y_\beta'' = 0. \qquad (4.3.2)$$

$$\eta_0'' + h^2\eta_0 = h, \qquad \eta_1'' + h^2\eta_1 = -h(1-2h\eta_0). \qquad (4.3.4)$$

*Quadrupole*

$$x_\beta'' + k_1\cdot\left(1-\delta+\delta^2\right)\cdot x_\beta = 0, \qquad y_\beta'' - k_1(\delta)y_\beta = 0. \qquad (4.3.5)$$

$$\eta_0'' + k_1\eta_0 = 0, \qquad \eta_1'' + k_1\eta_1 = k_1\eta_0. \qquad (4.3.6)$$

*Sextupole*

$$x_\beta'' = -\frac{1}{2}k_2\left(1-\delta+\delta^2\right)\left(x_\beta^2 - y_\beta^2\right) - k_2\left[\eta_0\delta - (\eta_0-\eta_1)\delta^2\right]x_\beta, \qquad (4.3.7)$$

$$y_\beta'' = k_2\left(1-\delta+\delta^2\right)x_\beta y_\beta + k_2\left[\eta_0\delta - (\eta_0-\eta_1)\delta^2\right]y_\beta.$$

$$\eta_0'' = 0, \qquad \eta_1'' = -\frac{k_2}{2}\eta_0^2. \qquad (4.3.8)$$

Let us neglect weak focusing in dipoles. The main terms in the first and second order chromaticity compensation could be described in following conditions:

$$\delta: \sum_Q (k_1L_Q)\beta_{xQ} - \sum_S (k_2L_S)\eta_0\beta_{xS} = 0, \qquad \sum_Q (k_1L_Q)\beta_{yQ} - \sum_S (k_2L_S)\eta_0\beta_{yS} = 0, \qquad (4.3.9)$$

$$\delta^2: \sum_Q (k_1L_Q)\beta_{xQ} - \sum_S (k_2L_S)(\eta_0-\eta_1)\beta_{xS} = 0, \qquad \sum_Q (k_1L_Q)\beta_{yQ} - \sum_S (k_2L_S)(\eta_0-\eta_1)\beta_{yS} = 0,$$

where subscripts Q and S denote summation over quadrupoles and sextupoles respectfully.

As it could be seen from the obtained equations if one will cancel the first order chromaticity (this is a goal of chromaticity correction section), and at that one is able to place strong sextupoles where $\eta_1 = 0$, then second order chromaticity will consequently cancel.

## 5. SIMULATION EXAMPLE

As an example we took the part of the final focus section containing vertical chromaticity correction section of the cτ factory project being developed in Novosibirsk [6]. Optical functions and main elements are shown on Fig. 5.1.

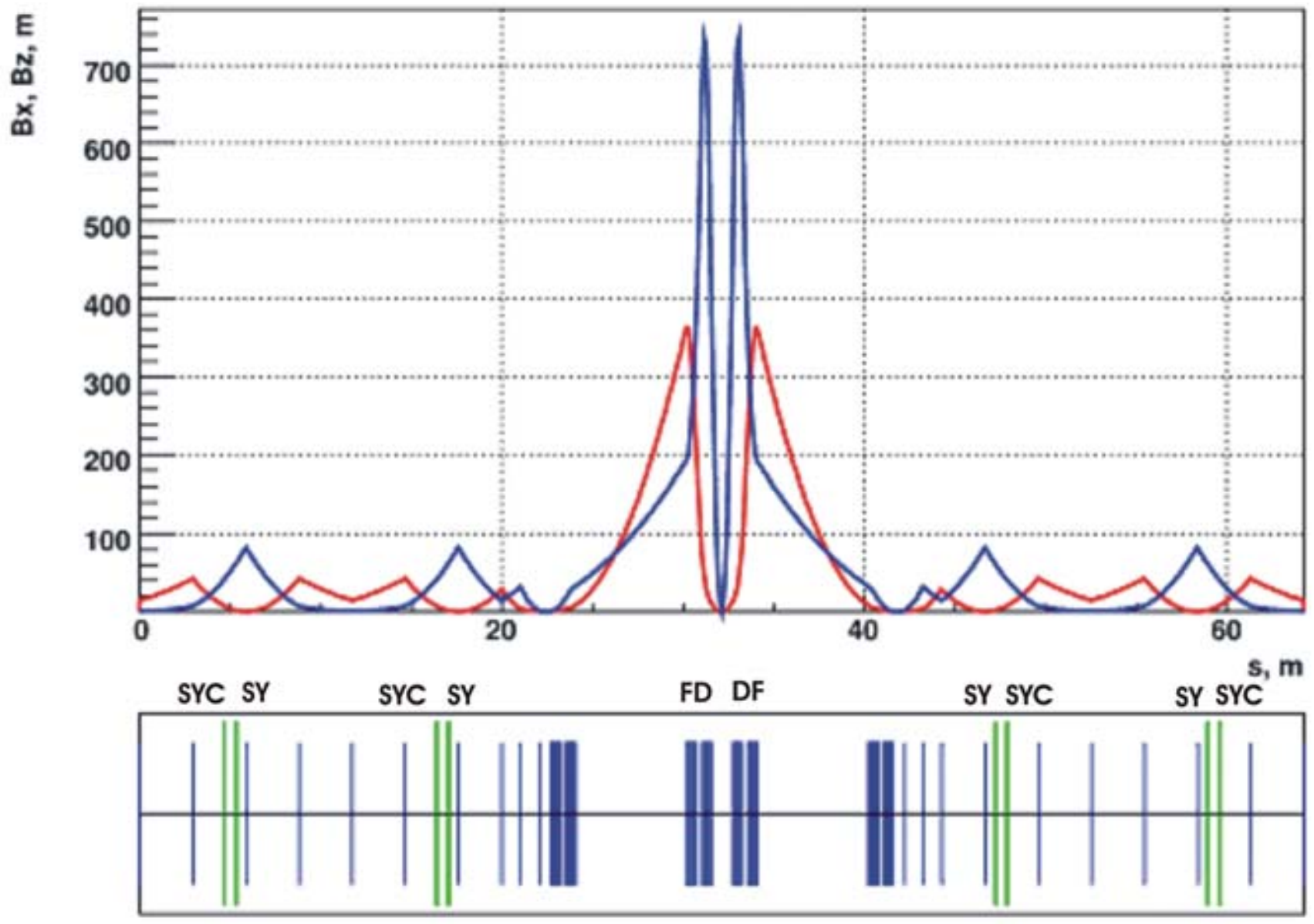

Fig.5.1 Optical functions (blue – vertical, red – horizontal) of final focus system

On the plot *FD* and *DF* denote final doublet, *SY* – the main pair of sextupoles, *SYC* – correcting pair of sextupoles.

In our case correcting sextupoles could be placed only after main ones (IP is a starting point), since the place before is occupied by quadrupoles. In other words this is the case described in previous paragraph when $S_1 = SY$ is the main pair, and $S_2 = SYC$ is correcting one.

Let us suggest the following procedure of aperture increase with the help of correcting sextupoles:

(1) We place correcting sextupoles *SYC* with arbitrary initial distance from the main ones and choose their strength in accordance with results shown on Fig.4.2.2. For example, at $\Delta L = L$ (where $L$ is sextupole length i.e. $k = 1$) for $B_1 = 0$ the strength of $SYC$ should be –14% from the main one, and –8% for $B_2 = 0$.

(2) We find borders of dynamic aperture for both solutions.

(3) We numerically find strength of $SYC$ providing the maximum of the dynamic aperture (in our case the merit is a maximum area surrounded by the DA curve).

(4) We shift the correcting sextupole in longitudinal direction and repeat previous steps in order to find maximum aperture in dependence on the distance between correcting and main sextupoles.

Example of application of described algorithm for point $k=1$ is shown on Fig.5.2. As one can see cancellation of aberrations in momentum part of transformation (3.4.1) ($B_2=0$) with relative strength of the correcting sextupole –8% is increasing linear size of the dynamic aperture two times in comparison with initial one. Cancelation of aberrations in coordinate part ($B_1=0$) with correcting sextupole strength –14% is more effective, increasing aperture 4-5 times.

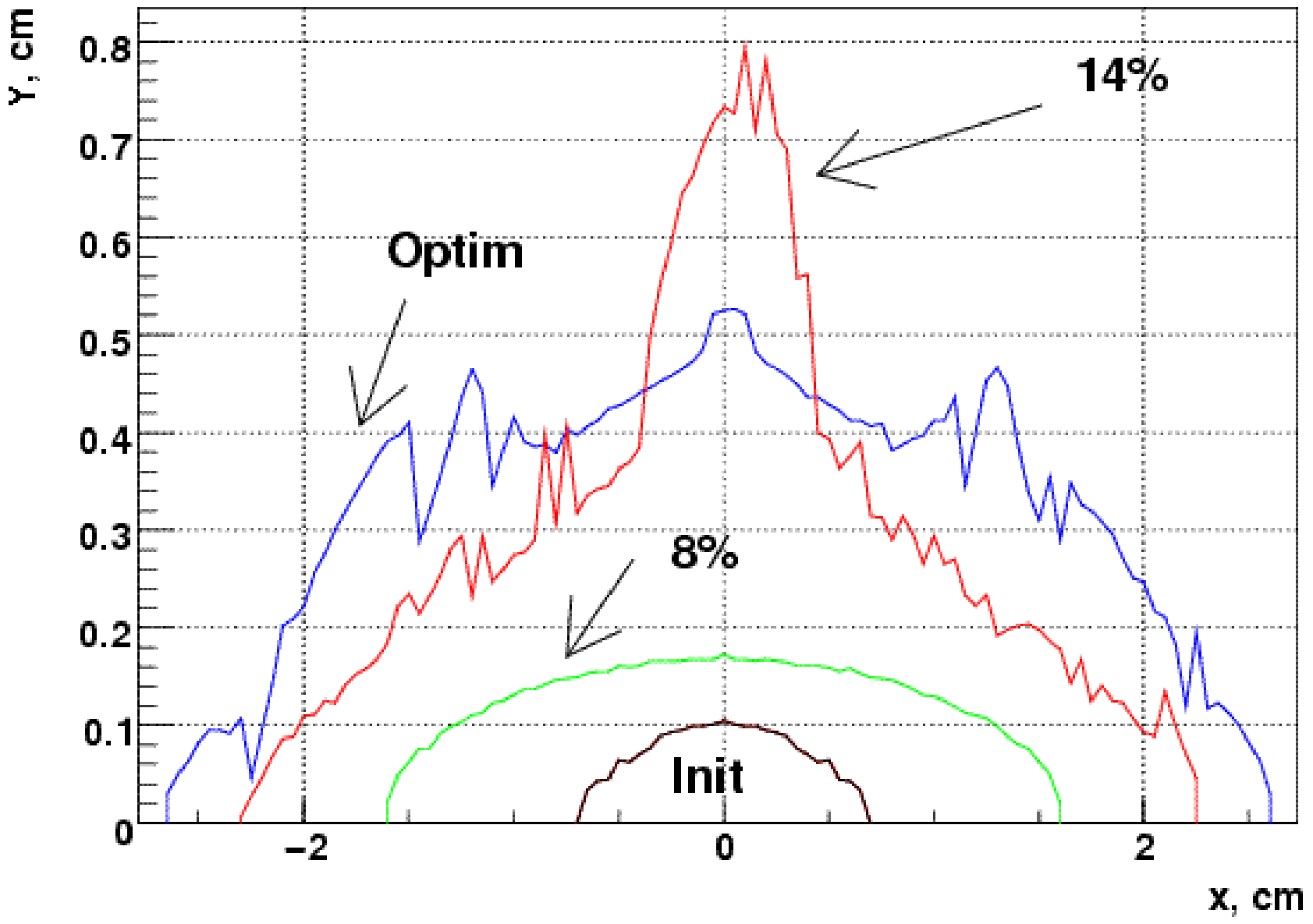

Fig.5.2 DA increasing with the help of correcting sextupoles. Initial aperture is shown in black

Additional numerical optimization gives slightly bigger aperture with the correcting sextupoles strength –10% from the main ones.

The values of relative correcting sextupole strength found from simulation and giving the maximum aperture, in dependence on the distance between the sextupoles is shown on Fig.5.3. It is seen that at small distances numerical results are in a good agreement with theoretical predictions, being somewhere in between case of $B_1=0$ and $B_2=0$.

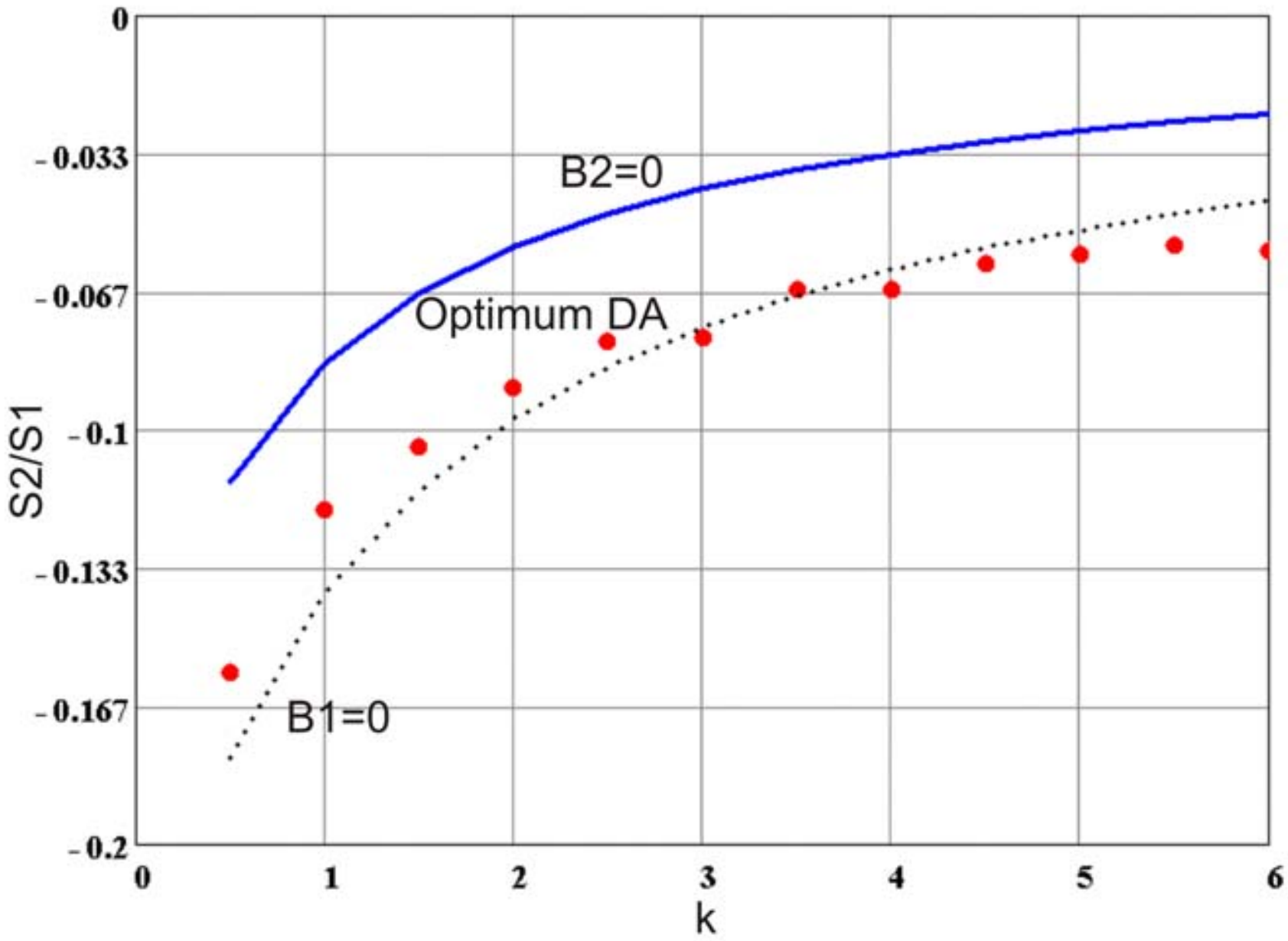

Fig.5.3 Numerically found correcting sextupole strength (dots) in comparison with analytical predictions

Agreement becomes worth with distance increase, probably because of defects of discussed approximation and increase of influence of higher order aberrations.

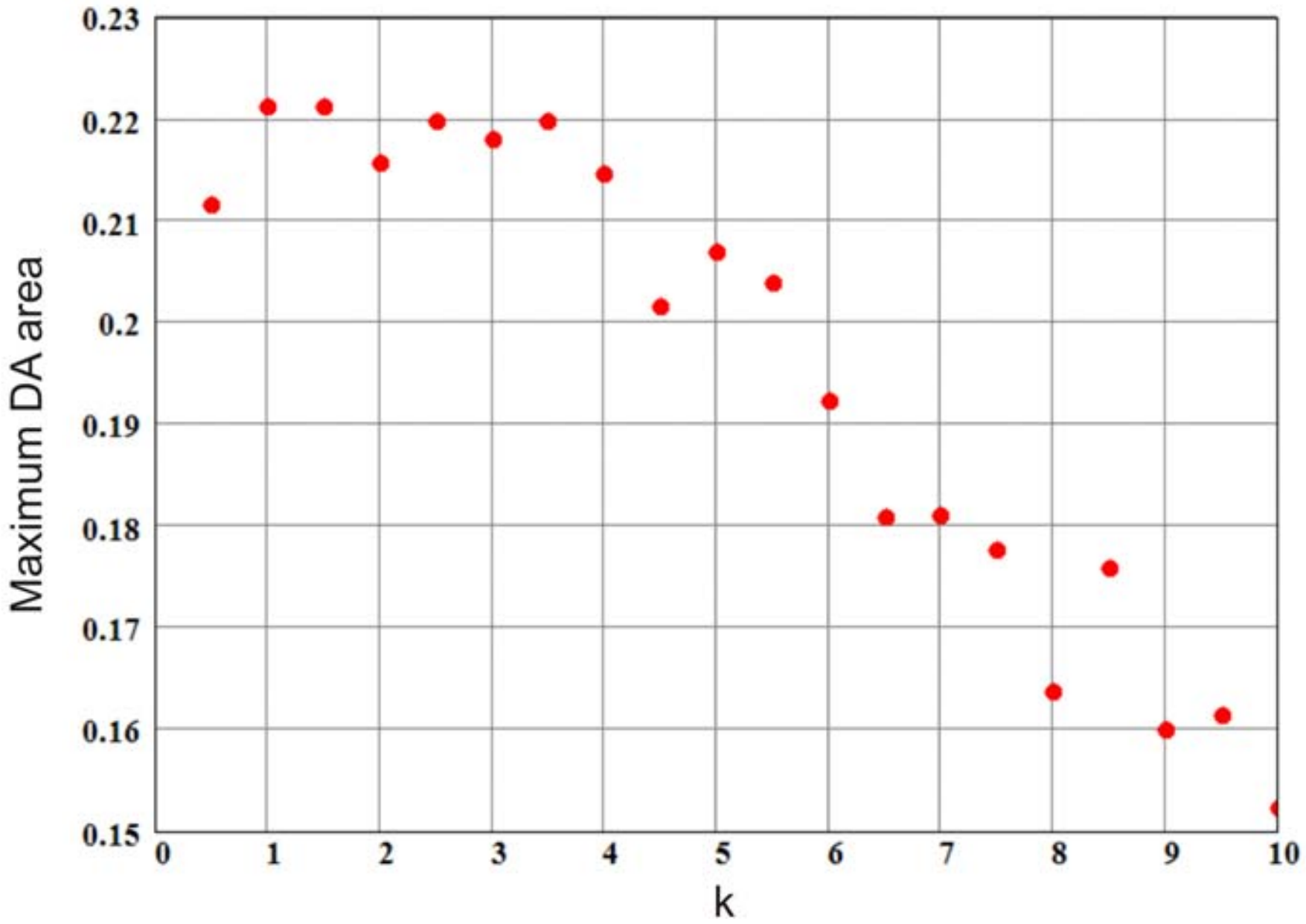

Fig.5.4 Optimized dynamic aperture in dependence on distance between the main and correcting sextupoles

Note, numerically optimized dynamic aperture decreases with increase of distance between sextupoles Fig.5.4. Therefore our recommendations for increase of dynamic aperture deteriorated due to non-zero length of the sextupoles of chromaticity correction section are:

- Add additional pair of sextupoles at the distance of ~1 to 4 sextupole length from the main one. It is assumed that between both pairs of sextupoles the transport matrix is negative identity.

- Choose the strengths of correcting sextupoles to cancel coefficients $B_1 = 0$ or $B_2 = 0$ (probably, $B_1 = 0$ is more preferable). With equal lengths of the main and correcting sextupoles, the strength of correcting ones are small, about $\leq 10\%$ from the main ones.

- Numerically optimize (if necessary) strength of correcting sextupoles in order to obtain maximum transverse aperture, which by experience is ~4-8 times bigger than initial uncorrected one.

- Providing $\eta_1 = 0$ at the place of strong sextupoles (by optics outside the chromatic section), than such chromaticity correction section should not significantly decrease the energy dynamic aperture (bandwidth).

## 6. CONCLUSION

Chromaticity correction section of the final focus for circular collider consists of two sextupoles, separated by *–I* map. Theoretically, such a structure cancels all geometrical aberrations of all orders. However, consideration of the finite sextupole length leads to cancelation of quadratic aberrations only, at that time aberrations of higher orders appear and sharply deteriorate dy-namic aperture. Significant increase of the aperture is provided with installation of additional weak sextupole pair, which will minimize aberrations of higher order.

This method could be used in other case when it is necessary to compensate effect of non-zero sextupole length, for example for "crab" sextupoles providing turn of the beam waist in the interaction point of the collider working in crab-waist scheme [7].